\documentstyle[prl,aps,multicol,epsf]{revtex}

\newcommand{\ve}{\varepsilon}
\newcommand{\bm}[1]{\mbox{\boldmath$#1$}}
\begin{document}
\draft

\title{Rashba precession in quantum wires}
\author{Wolfgang H\"ausler
\thanks{Email: haeusler@physnet.uni-hamburg.de}
}
\address{Physikalisches Institut,
Universit\"at Freiburg,
Hermann-Herder-Str. 3,
79104 Freiburg, Germany}
\date{Received:\hspace*{3cm}}
\maketitle
\begin{abstract}
The length over which electron spins reverse direction due to
the Rashba effect when injected with an initial polarization
along the axes of a quantum wire is investigated theoretically.
A soft wall confinement of the wire renormalizes the spin-orbit
parameter (and the effective mass) stronger than hard walls.
Electron-electron interactions enhance the Rashba precession
while evidence is found that the coupling between transport
channels may suppress it.
\end{abstract}

\pacs{Keywords: Rashba precession, spin splitting, spin transport,
Tomonaga-Luttinger model}
\begin{multicols}{2}
\narrowtext

Spin transport has regained considerable interest in recent
years, partly because of the prospective for qualitatively novel
electronic devices and partly since fundamentally new mesoscopic
quantum coherence phenomena could be investigated in the
magnetic degree of electronic freedom. A prime example is the
spin transistor proposed by Datta and Das \cite{dattadas}.

Here we investigate how Rashba spin precession \cite{rashba} is
affected by many body effects \cite{raikh} as expected to be
important in semiconducting devices. We consider electrons
moving along a one-dimensional (1D) quantum `wave guide'. Also
the cases of two interacting channels and different transversal
confinement potentials are addressed. Of prime interest will be
the length $\lambda_{\rm R}$ over which spins initially
polarized along the channel axes reverse spin-direction while
moving along the channel.

In a 2D layer spin splitting is a consequence
of spin-orbit coupling
\begin{equation}\label{hso}
H^{\rm so}=\alpha(\sigma_xp_z-\sigma_zp_x)
\end{equation}
proportional to the Rashba parameter $\alpha$ and thus to the
intrinsic or by means of gates externally applied electric field
in $y$--direction, perpendicular to the layer \cite{rashba}. $H^{\rm
so}$ is also proportional to the momentum of spins $(p_x,p_z)$
in the `active' $x-z$--plane, $\sigma_{x,z}$ are Pauli matrices.
Precession then occurs on the length scale $|k_+-k_-|^{-1}$ of
the spin split wave numbers $k_{\pm}$ at the Fermi energy. If the
effective mass approximation $\ve(k)=k^2/2m$ is applicable, as
for many semiconductors, the Rashba length
\begin{equation}\label{lambdar}
2\pi/|k_+-k_-|=\pi/m\alpha
\end{equation}
does {\em not} depend on carrier density or Fermi energy.

With electrons being confined in $z$--direction, to zero'th order
in $H^{\rm so}$ this picture carries over to 1D. Only to
higher order $H^{\rm so}$ will weakly mix different transport
channels of the wave guide which up to O$(\alpha^5)$ can be
accounted for by renormalizing $\alpha\to\alpha^*$ in
$\ve_{\pm}(k)$ and, within the effective mass approximation, by
$m\to m^*$. In the latter case the Rashba length modifies
according to $(m\alpha)^{-1}\to(m^*\alpha^*)^{-1}$. For a
quantitative estimate of both renormalizations the intra-subband
eigenfunctions
\begin{equation}\label{psins}
\psi_{kns}(x,z)={\rm e}^{{\rm i}kx}\phi_n(z)
(\cos(m\alpha z)|s\rangle+{\rm i}\,\sin(m\alpha z)|\!-\!\!s\rangle)
\end{equation}
are needed, which are plane waves of momentum $k$ along the wave
guide and, without inter-subband scattering, slightly modified
subband states $\:\phi_n\:$ (subband index $n$) in
$z$--direction, spin polarized $s=\pm$ on the axes at $z=0$.
For a harmonic confinement (subband energy $\omega_0$), as
relevant for example for samples defined by gating \cite{tarucha},
the perturbative estimate yields $\:\alpha^*=\alpha(1-\eta)\:$
and $\:m^*=m(1+8\eta^2)\:$ in the ground subband. The
dimensionless parameter $\eta=(wm\alpha/2)^2$ measures the bare
value of the spin-orbit strength $\alpha$ in (\ref{hso}) by
comparing the wave guide width $w=2/\sqrt{m\omega_0}$ with the
length $(m\alpha)^{-1}$. For a hard wall confinement on the
other hand (again of width $w$), as possibly more relevant for
wires fabricated by the cleaved edge technique \cite{yacoby},
the renormalizations become
$\:\alpha^*=\alpha(1-(1/6-1/\pi^2)\eta)\:$ and
$\:m^*=m(1+3(4/3\pi)^6\eta^2)\:$, i.e.\ they are significantly
reduced compared to the soft wall case since $1/6-1/\pi^2\approx
0.065$ and $3(4/3\pi)^6/8\approx 0.002$.

The main idea of the transistor operation \cite{dattadas} is to vary
the strength of the electric field and thereby $\pi/m\alpha$.
This is achieved by a charging a gate parallel to the layer.
Unless carefully compensated by a second back gate \cite{grundler},
however, the carrier density in the structure will change at the
same time. Ignoring interactions this would be unimportant when
$\ve(k)=k^2/2m^*$ but the interaction strength depends, even
quite sensitively, on carrier density through the $r_{\rm
s}$--parameter. In order to investigate possible consequences
for the Rashba precession we employ the Tomonaga-Luttinger (TL)
model \cite{tl} as the most precise low energy description of 1D
metals. Here we shall not focus on various very characteristic
power laws predicted by this model, also regarding spin
properties \cite{egger-epl} coming from the charge sector, but rather
focus on the question how interactions influence the length
$\lambda_{\rm R}$.

As a second striking property the TL-model predicts charge--spin
separation which, interestingly and contrary to statements in
the literature \cite{egger-epl,moroz}, is {\em not} spoiled in
semiconducting quantum wires unless spin-orbit coupling is not
exceedingly strong $\eta\gtrsim 1$ to devastate the effective
mass description. On the other hand, for non-quadratic
dispersion relations, charge--spin separation is in general
destroyed with $H^{\rm so}$. An example are carbon nanotubes
where $v_\pm=v_{\mathrm{F}}\pm\alpha$ with $\alpha$ originating
in this case from the curved surface \cite{cylinder} so that $v_+\ne
v_-$ spoils charge--spin separation.

A realistic interaction in 1D
\begin{equation}\label{interaction}
V(x)=\frac{e^2}{\kappa}\left(\frac{1}{\sqrt{x^2+w^2}}-
\frac{1}{\sqrt{x^2+w^2+4R^2}}\right)
\end{equation}
is specified by the wire width $w$ and a cut-off at large
distance $R$ due to screening by the nearest metals, such as the
gates that help defining the quantum channel. Since spin
properties involve finite momenta $q$ of the Fourier transformed
interaction $\hat V(q)$ (and contrary to charge properties not
the dominant long wave length limit $\hat V(q\to 0)$) the
interaction range $R$ plays only a minor r\^ole if $(R/w,k_{\rm
F}R)\gg 1$. In carbon nanotubes, for example, interactions will
alter the spin velocity only weakly, in marked contrast to the
charge velocity \cite{mceuen99a}.

How to include the Rashba term in the TL-model~? In previous
work \cite{egger-epl,moroz} the Fermi velocities $v_+$ and $v_-$ have
simply been put to different values, which for $\ve(k)=k^2/2m^*$
does not describe the leading effect of $H^{\rm so}$. Rather
{\em both} velocities change slightly but obey $v_+=v_-$ and
charge--spin separation. In effective mass systems $H^{\rm so}$
acts solely in the spin sector of the corresponding TL low
energy model (of length $L$), where the topological term
\begin{equation}\label{bosontop}
\frac{\pi}{4L}\left(v_{\mbox{\tiny N}}N_{\sigma}^2+
v_{\mbox{\tiny J}}J_{\sigma}^2\right)-m^*\alpha^*v_{\mathrm{F}}
J_{\sigma}
\end{equation}
is most relevant for the following. $N_{\sigma}$ and
$J_{\sigma}$ denote the usual currents of velocities
$v_{\mbox{\tiny N/J}}$ where with Coulomb repulsion
$(v_{\mbox{\tiny N}},v_{\mbox{\tiny J}})\ne v_{\mathrm{F}}$
\cite{tl}. In strictly spin isotropic systems $v_{\mbox{\tiny
N}}=v_{\mbox{\tiny J}}$. Since we expect this isotropy to be
broken only weakly, both of these velocities should be similar
in magnitude and also similar to the velocity $v_\sigma$ of spin
wave propagation. This latter quantity has been determined
recently by extensive quantum Monte--Carlo simulations
\cite{creffield}. With increasing interaction strength,
equivalent to a decreasing carrier density,
$v_\sigma/v_{\mathrm{F}}$ was found to decrease, even
below 0.5 when accounting for parameters of existing
quantum wires \cite{yacoby}.

Many quantities of interest can be calculated exactly using
(\ref{bosontop}). In particular it can be shown \cite{rapid}
that spins polarized in $x$--direction along the wire precess
now over a length that acquires a factor $v_{\mbox{\tiny J}}/
v_{\mathrm{F}}$ compared to the length of Eq.~(\ref{lambdar}).
With $v_{\mbox{\tiny J}}=v_\sigma$ we conclude from
\cite{creffield} that this length {\em decreases} with
increasing interaction strength. A similar conclusion has been
drawn for two-dimensional electrons after treating the
interaction perturbatively \cite{raikh}. This trend is opposite
to what is expected for almost linear single electron
dispersions (such as narrow gap semiconductors) but agrees with
experimental observations \cite{experiments}.

For very large and very small $k_{\rm F}w$ asymptotic
expressions can be given for $v_{\sigma}$ which result in the
following dependencies for the Rashba length:
\[
\lambda_{\rm R}\sim\frac{\pi}{m^*\alpha^*}
[1-f(2k_{\rm F}w)]\;\;\mbox{if}\;\;f(2k_{\rm F}w)<\frac{1}{2\pi}
\]
at large $k_{\rm F}$, $\:f(x):=\sqrt{\frac{2}{\pi}}\frac{w}{a_{\rm B}}
x^{-3/2}\:{\rm e}^{-x}\:$, and
\[
\lambda_{\rm R}\sim\frac{\pi^2}{3m^*\alpha^*}\frac{a_{\rm B}}{w}
\frac{2k_{\rm F}w}{\ln 2R/w}\;\;\mbox{if}\;\;k_{\rm F}\ll 1/R\;,
\]
$a_{\rm B}$ is the Bohr radius. This latter estimate follows
from the conjecture \cite{whlkahm} that the system falls into the
universality class of the Hubbard model at small fillings in
which case the interaction range $R$ becomes relevant.

Now we turn to the case of two occupied channels in the quantum
wire. Then the renormalizations $\alpha\to\alpha_j$ and $m\to
m_j$ due to $H^{\rm so}$ acquire a channel index $j$; for
example, in a harmonic confining potential $\alpha_2$ of the
upper channel renormalizes by a factor of 3 more than
$\alpha_1$.

Between the channels acts the same microscopic (screened Coulomb
type) interaction (\ref{interaction}) as within each channel. Up
to now no theory is available to determine spin velocities in
multi-channel situations and we would like to sketch a possible
solution here \cite{details}. We assume
1) sufficiently different carrier densities in both channels to
conserve the TL--phase \cite{starykh}
(otherwise spin gaps could spoil the Rashba precession), and 2)
charge--spin separation.  The Fermi momentum in the upper
channel may be taken as
$\:k_2=\sqrt{\frac{m_2}{m_1}k_1^2-2m_2\omega_0}\:$
\cite{density} where $\omega_0$ is the subband spacing and
$m_{2/1}$ are the effective masses of the upper/lower channel
after renormalization through $H^{\rm so}$.

We expect two spin velocities $v_{\sigma 1}$ and $v_{\sigma 2}$
which both will be relevant for the Rashba precession, similar
to the single channel case. They can be obtained from the
microscopic interaction through generalized susceptibilities
\begin{equation}\label{chargesusc}
(\bm{v}_{\mbox{\tiny N}})_{ij}=\frac{L}{\pi}\frac{\partial^2E_0}
{\partial N_i\partial N_j}
\end{equation}
and
\begin{equation}\label{currentsusc}
(\bm{v}_{\mbox{\tiny J}})_{ij}=\frac{L}{\pi}\frac{\partial^2E_0}
{\partial J_i\partial J_j}\;,
\end{equation}
when suitably generalizing the relation
$\:v_{\sigma}=\sqrt{v_{\mbox{\tiny N}}v_{\mbox{\tiny J}}}\:$
observed in one channel. Here, $N_j$ and $J_j$ are the currents
(\ref{bosontop}) in channel $j$.

How can we obtain $\bm{v}_{\mbox{\tiny N}}$ and
$\bm{v}_{\mbox{\tiny J}}$ for a given microscopic interaction~?
In principle we know already from the single channel case how
notoriously difficult spin velocities are to evaluate
\cite{creffield}. Leading corrections at not too small carrier
densities to the ground state energy $E_0/L$ per length can be
obtained perturbatively. Fock-exchange terms $\:\sim\hat
V(k_i\pm k_j)\:$ have to be included adequately for spin
properties, which goes beyond the RPA approximation mostly used
for estimating charge velocities \cite{joynt} since the $\hat
V(q=0)$ contribution drops out here.

The result is
\begin{equation}\label{chargepert}
\bm{v}_{\mbox{\tiny N}}=\left(\matrix{
v_1-\frac{\hat V(2k_1)}{\pi}+\hat V_-&-\hat V_+\cr
-\hat V_+&v_2-\frac{\hat V(2k_2)}{\pi}+\hat V_-
}\right)
\end{equation}
and
\begin{equation}\label{currentpert}
\bm{v}_{\mbox{\tiny J}}=\left(\matrix{
v_1+\hat V_-&-\hat V_-\cr
-\hat V_-&v_2+\hat V_-
}\right)\;.
\end{equation}
Here, $v_j$ are the bare Fermi velocities and $\hat V_{\pm}=(\hat
V(k_1-k_2)\pm\hat V(k_1+k_2))/2\pi$. Both limits, $k_2\to 0$
(almost empty upper channel) and $k_2\to k_1$ (limit of
equivalent subbands), indicate instabilities in
$\bm{v}_{\mbox{\tiny N}}$ and $\bm{v}_{\mbox{\tiny J}}$ when
$\hat V(q=0)>\pi v_j$ (long range interactions, note that
$V(q)$ decreases with increasing $q$ for any realistic electron-electron
interaction). These are precursors of the Cooper or charge
density wave instabilities \cite{starykh}, that occur in
repulsively interacting two-channel systems near the threshold
for opening the second channel or near equal carrier densities
in both channels, respectively. We also see in
(\ref{currentpert}) that $\bm{v}_{\mbox{\tiny J}}$ leaves the
Galilei mode (1,1) independent of the interaction. For
$v_1=v_2$ this mode is an eigenmode so that perturbation theory
pretends Galilei invariance of the spin sector, similar as in
the single channel case \cite{whlkahm}. Then perturbative
renormalization of the spin velocities is solely due to
$\bm{v}_{\mbox{\tiny N}}$.

\begin{figure}
\epsfxsize=0.8\columnwidth
\centerline{\epsffile{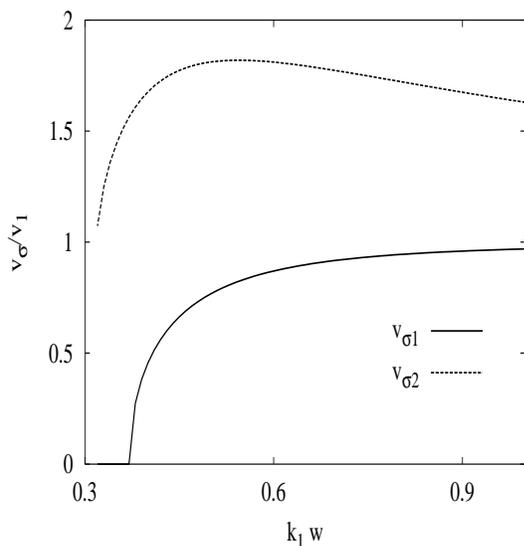}}
\caption[]{
Spin velocities in a two channel quantum wire in units of the larger
of the Fermi velocities $v_1$, versus $k_1w$ in perturbation theory.
}
\label{vsm}
\end{figure}

Resulting spin velocities are shown in Fig.~\ref{vsm}. One of
the velocities, $v_{\sigma 1}$, is decreasing with decreasing
carrier density $2k_1/\pi$, similar as in the single channel
case \cite{creffield}. More strikingly, $v_{\sigma 2}$ first
increases and exceeds the larger of the two Fermi velocities
$v_1$. Then according to Eq.~(\ref{lambdar}) we have evidence
for a {\em suppression} of the Rashba precession as a result of
inter-channel coupling by electron-electron interaction.

In conclusion, we have discussed how a hard or soft
confining potential, and the intra- and inter-channel
interaction affect the Rashba precession along a quantum wire.

I would like to thank Hermann Grabert for continuous support.

\end{multicols}
\end{document}